\documentstyle[12pt]{article}

\begin{document}

\topmargin 0pt
\oddsidemargin 5mm
\def\bbox{{\,\lower0.9pt\vbox{\hrule \hbox{\vrule height 0.2 cm
\hskip 0.2 cm \vrule height 0.2 cm}\hrule}\,}}

\newcommand{\EQ}{\begin{equation}}
\newcommand{\EN}{\end{equation}}
\newcommand{\bea}{\begin{eqnarray}}
\newcommand{\ena}{\end{eqnarray}}
\newcommand{\hs}[1]{\hspace{#1 mm}}
\newcommand{\shalf}{\frac{1}{2}}
\newcommand{\tri}{{\small $\triangle$}}
\newcommand{\dz}{\frac{dz}{2\pi i}}
\newcommand{\ra}{\rangle}
\newcommand{\lan}{\langle}
\newcommand{\nn}{\nonumber \\}
\def\a{\alpha}
\def\b{\beta}
\def\g{\gamma}
\def\G{\Gamma}
\def\d{\delta}
\def\D{\Delta}
\def\e{\epsilon}
\def\ve{\varepsilon}
\def\z{\zeta}
\def\t{\theta}
\def\vt{\vartheta}
\def\r{\rho}
\def\vr{\varrho}
\def\k{\kappa}
\def\l{\lambda}
\def\L{\Lambda}
\def\m{\mu}
\def\n{\nu}
\def\o{\omega}
\def\O{\Omega}
\def\s{\sigma}
\def\vs{\varsigma}
\def\S{\Sigma}
\def\vphi{\varphi}
\def\av#1{\langle#1\rangle}
\def\pa{\partial}
\def\na{\nabla}
\def\hg{\hat g}
\def\un{\underline}
\def\ov{\overline}
\def\cF{{\cal F}}
\def\Hsl{H \hskip-8pt /}
\def\Fsl{F \hskip-6pt /}
\def\cFsl{\cF \hskip-5pt /}
\def\ksl{k \hskip-6pt /}
\def\pasl{\pa \hskip-6pt /}
\def\tr{{\rm tr}}
\def\tcF{{\tilde{\cal F}}}
\def\tg{{\tilde g}}
\def\cLNSI{{\cal L}^{\rm NS}_{\rm I}}
\def\bcLNSI{{\bar\cLNSI}}

\newcommand{\AP}[1]{Ann.\ Phys.\ {\bf #1}}
\newcommand{\NP}[1]{Nucl.\ Phys.\ {\bf #1}}
\newcommand{\PL}[1]{Phys.\ Lett.\ {\bf #1}}
\newcommand{\CMP}[1]{Comm.\ Math.\ Phys.\ {\bf #1}}
\newcommand{\PR}[1]{Phys.\ Rev.\ {\bf #1}}
\newcommand{\PRL}[1]{Phys.\ Rev.\ Lett.\ {\bf #1}}
\newcommand{\PTP}[1]{Prog.\ Theor.\ Phys.\ {\bf #1}}
\newcommand{\PTPS}[1]{Prog.\ Theor.\ Phys.\ Suppl.\ {\bf #1}}
\newcommand{\MPL}[1]{Mod.\ Phys.\ Lett.\ {\bf #1}}
\newcommand{\IJMP}[1]{Int.\ Jour.\ Mod.\ Phys.\ {\bf #1}}
\newcommand{\CQG}[1]{Class.\ Quant.\ Grav.\  {\bf #1}}
\newcommand{\PRep}[1]{Phys.\ Rep.\ {\bf #1}}
\newcommand{\RMP}[1]{Rev.\ Mod.\ Phys.{\bf #1}}

\begin{titlepage}
\setcounter{page}{0}

\begin{flushright}
COLO-HEP-371 \\
hep-th/9607011 \\
June 1996
\end{flushright}

\vspace{5 mm}
\begin{center}
{\large A note on brane tension and M-theory }
\vspace{10 mm}

{\large S. P. de Alwis\footnote{e-mail:  
dealwis@gopika.colorado.edu}}\\
{\em Department of Physics, Box 390,
University of Colorado, Boulder, CO 80309}\\
\vspace{5 mm}
\end{center}
\vspace{10 mm}

\centerline{{\bf{Abstract}}}
We point out that some M-theory results for brane tension, can be  
derived from
Polchinski's formula for D-brane tension. We also argue that this  
formula
determines gravitational and gauge couplings in the low energy but   
quantum
exact  effective action.

\end{titlepage}
\newpage
\renewcommand{\thefootnote}{\arabic{footnote}}
\setcounter{footnote}{0}

\setcounter{equation}{0}

Polchinski's formula for the tensions of D-branes \cite{jp} states  
that the
tension $T_p$ of a D (p)-brane is given by,

\begin{equation}\label{jp}
T_p^2={2\pi\over 2\k^2} (4\pi^2\a ')^{3-p},
\end{equation}
where $1/2\pi\a'$ is the fundamental string tension and $2\k^2$ is  
($16\pi$
times) Newton's constant in  10 dimensions. Now since type IIA string  
theory
has only even $p$  and type IIB only
odd $p$ D-branes one might wonder whether the even and odd series of  
the above
formula should be treated separately.
However since the two series are related
by T-duality one should take this to be one series with one ten  
dimensional
gravitational coupling. This is clarified by recent work of Green,  
Hull, and
Townsend \cite{ght}  who
find\footnote{It should be noted that one needs to replace the  
non-standard $\a
'$ in \cite{ght}    according to $\alpha '\rightarrow 4\pi^2\a '$ to  
get the
usual definition.} from T-duality that
\begin{equation}\label{ght}
T_{p-1} =2\pi\a '^{1/2}T_{p}
\end{equation}
in agreement with (\ref{jp}). It is curious that this relation  
implies that the
(p-1)-brane tension
can be obtained from that of the p-brane by compactifying on a circle  
with the
self dual (physical) radius.

 Note  that (\ref{ght}) implies $T_p=(4\pi^2\a ')^{3-p}T_{6-p}. $  
Substituting
in the brane quantization rule \cite{dirac}\footnote{For a review see
\cite{duff}.}
\begin{equation}\label{}
2\k^2 T_pT_{6-p}=2\pi n,~~n~\e ~{\bf Z}
\end{equation}
we get, for $n=1$, equation (\ref{jp}). In other words the T-duality  
result
plus the quantization condition gives Polchinski's formula for  
D-brane
tensions.

Now when $p=1$ we have the D-string of type IIB theory and by  
$SL(2,Z)$ duality
\cite{js}\footnote{Actually as shown in \cite{schmid} \cite{ds} one  
need not
{\it assume} $SL(2,Z)$ to get this.}, it
has tension  $g^{-1}(2\pi\a ')^{-1}$,  where the second factor is the
fundamental string tension, and the relation gives a definition of  
the
effective string coupling which for  weak coupling takes the form
$g=<e^{\phi}>+\ldots$ where $\phi$ is the dilaton. Hence we have from
(\ref{jp}) a relation between $\k$
and $\a '$,

\begin{equation}\label{kappa}
2\k^2=(2\pi )^3 (2\pi\a ')^4g^2
\end{equation}

Let us now discuss the correspondence with M-theory results. Putting
$T=1/2\pi\a '$  the fundamental string tension, we find from  
(\ref{jp}) and
(\ref{kappa}).

\begin{equation}\label{IIB}
{T^2\over T_3}={1\over (2\pi\a  
')^2}{\sqrt{2\k^2}\over\sqrt{2\pi}}=2\pi g
\end{equation}
Similarly one may obtain $T^3/T_5 = (2\pi )^2g$.  But these are  
precisely the
formulae obtained by  Schwarz
 \cite{js2}\footnote{Note that we have expressed all tensions in the  
string
metric hence the extra factor $g$ as compared with \cite{js2}  
equation (6).
Also we have set
the axion field to zero for simplicity.}
 using M-theory. Here they have been obtained purely from string  
theory. This
is
not so surprising since they relate quantities that are all in type  
IIB.

 Schwarz \cite{js2} has also obtained using M-theory arguments, the  
relation
\begin{equation}\label{M}
{T_5^M\over (T_2^M)^2}={1\over 2\pi},
\end{equation}
 between the  M-two-brane and the M-5-brane tension\footnote {It   
should be
stressed that the ratios on the left hand sides of
(\ref{IIB}, \ref{M}) are dimensionless
and hence independent of the metric in which the two tensions are  
defined.}.
 We note in passing that a  similar relation (\ref{M}) had been  
obtained
earlier by Duff et al
\cite{dlm} by using M-theory quantization conditions   at membrane  
and at 11
dimensional effective supergravity action level, but with the right  
hand side
being twice that given above\footnote{This is similar to the factor  
two
discrepancy found by Polchinski \cite{jp} between his results and an  
effective
action calculation of  Harvey and Strominger \cite{hs} .}. In an  
appendix we
rederive
the Duff et al result and find that it actually agrees with (\ref{M})  
above
(and hence as shown below with the Polchinski quantization condition  
as well)
thus eliminating a potential conflict.
We will  obtain (\ref{M}) from  the
purely type IIA relation between the 2-brane and the 4-brane tensions  
which are
determined by
(\ref{jp}).

The M-theory metric is related to the type IIA one by (see for  
example
\cite{pt})

\begin{equation}\label{metric}
ds_M^2 =g^{4/3}dx_{11}^2+g^{-2/3}ds_{10}^2
\end{equation}
We adopt the convention that the coordinate range of the circular  
11th
dimension goes over
$0-2\pi\sqrt\a'$ where $1/2\pi\a' = T$ is now the IIA tension. This  
then serves
to define
the effective IIA coupling $g= <e^\phi+\ldots>$ in terms of the  
physical radius
of the circle $R_{11}=\sqrt{\a'} g^{2/3}$. In other words
$g^2=R_{11}^3T^{3/2}(2\pi )^{3/2}$ is {\it defined} (as in the IIB  
case) in
terms of physical quantities.  Let us denote the  induced string  
metric on the
world sheet  in IIA by $\g$, and the induced M theory metric on the  
2-brane
world volume by
$\g_{M2}$ . The IIA string is obtained from the M-2-brane by double  
dimensional
reduction, i.e. by wrapping one dimesion of the membrane around the  
circle of
M-theory.  Then from (\ref{metric}) we get,
$\sqrt\g_{M2}=\sqrt\g$ and hence $T_2^M 2\pi\sqrt \a'= T=1/2\pi\a '$  
giving

\begin{equation}\label{twoM}
T_2^M=1/(2\pi )^2 (\a ')^{3\over 2}
\end{equation}

Now  the relations between the  two induced metrics on
the 2-brane are given (using (\ref{metric})) by  $\sqrt{\g_{M2}}
=g^{-1}\sqrt{\g_2}$. The IIA 2-brane is obtained by simple  
dimensional
reduction, and one obtains \cite{pt} the relation
$T_2=g^{-1}T_2^M$. Using the expression $T^2_2 = (2\pi  
)^3\a'/2\kappa^2$ from
(\ref{jp}) and equation (\ref{twoM}) we get again (\ref{kappa})  
except that now
$g$ and $\a '$ are defined as above in the IIA theory.

The 4-brane is obtained from the M-theory 5-brane by double
dimensional reduction
\cite{pt}. Using again (\ref{metric}) we have for the world  volume  
densities,
$\sqrt\g_M^5 =g^{-1}\sqrt{\g^5}$ and using, for  consistency,  the  
coordinate
radius $\sqrt{\a '}$ as before, we have $T^A_4=g^{-1}T_5^M  
(2\pi\sqrt{\a '})$.
 Using the above relations between IIA tensions and M-tensions,  
Polchinski's
formula
 (\ref{jp}), and (\ref{kappa}),  we get
\begin{equation}\label{fiveM}
T_5^M=(2\pi )^{-5}\a '^{-3},
\end{equation}
{}From (\ref{twoM}) and (\ref{fiveM})  the M-theory relation  
(\ref{M}) is
recovered\footnote{A similar derivation of this result has been given  
by I.
Klebanov and A. Tseytlin \cite{kt}. I wish to thank Igor Klebanov for  
drawing
my attention to these papers. }.

To go further let us use  the quantization relation of M-theory,

\begin{equation}\label{Mquant}
2\k^2_{11}T_2^MT_5^M=2\pi n,~~n\e{\bf Z}
\end{equation}
 Then putting $n=1$ and using the results (\ref{twoM}) and  
(\ref{fiveM}), we
have,
\begin{equation}\label{keleven}
2\k_{11}^2=(2\pi )^{8}\a '^{9/2}.
\end{equation}
 This formula (which could also have been obtained by comparing the
gravitational actions of M-theory and IIA theory) should be  
interpreted as
fixing the string scale in terms of the
fundamental M-theory
scale, but for notational convenience we will continue to express  
everything in
terms of $\a '$.

It should be noted at this point that in comparing the expression
(\ref{keleven}) to the string
scale one has to be careful as to what metric the string scale is  
defined in.
We have defined
the fundamental string length to be $l_{string}\equiv\sqrt\a '$ in  
the string
metric. This means that in the M-theory metric this has the value  
(using the
conversion factor from the metric (\ref{metric}))   
$l_{string}^M=g^{-1/3}\sqrt
{\a '}$. Hence  we have from (\ref{keleven}) the expression,

\begin{equation}\label{}
l_{11}\equiv [2\k_{11}^2/(2\pi )^8]^{1\over 9}=g^{1/3}l^M_{string}.
\end{equation}
for the 11 dimensional Planck length. It should be noted also that  
our
definition of the coupling constant (in the discussion after equation
(\ref{metric})) is equivalent to the relation $g=R_{11}  
/l_{string}^M$. This
clarifies the
relation between our conventions and that of other authors \cite{kt},
\cite{kp}\footnote{I wish to thank Igor Klebanov for raising this  
issue.}.

Now a highly non-trivial M-theory result obtained by Horava and  
Witten
\cite{hw} is the value of the  (dimensionless) ratio  
$(2\k^2_{11})^2/\l^6
=(2\pi )^{-5}$, where $\l$ is the
$E_8\times E_8$ gauge coupling in the theory on
$R_{10}\times S_1\over Z_2$. To make a connection, let us calculate  
in type I
or I' theory,  making the plausible assumption  that  since these  
theories are
obtained by a certain projection on type II  theories the previous  
results on
the gravitational constant hold true.
To identify the gauge coupling let us look at the nine-brane action
\cite{rl},\cite{cjp} in the limit of flat space.
We have (see for example  \cite{cjp} equation (2.19))

\begin{equation}\label{}
T_9\tr\det [1+(2\pi\a ')F]^{1/2}\sim T_9{(2\pi\a ')^2\over 4}\tr F^2.
\end{equation}

Using the value of $T_9$ from the Polchinski formula (\ref{jp}) we  
have for the
gauge coupling
\begin{equation}\label{gauge}
\lambda^2=(2\pi )^{5/2}(2\pi\a ')\sqrt{2\k^2}=(2\pi )^7\a '^3g
\end{equation}

The upshot of these calculations is that we have the following  
equivalent forms
of low-energy effective string (M) theory (with gauge fields) in 10  
(11)
dimensions. Note that because of the BPS argument for the D-brane  
tensions it
is plausible that these are quantum effective actions with the actual  
physical
couplings at the string scale. i.e. they are low energy, but quantum  
exact,
effective actions, which can be used as the initial condition for the  
RG
evolution (after compactification) down to low energies. Even at  
strong
coupling the form of the low energy effective action is determined by  
general
covariance and gauge invariance, and our claim is
that the D-brane formula effectively determines the values of the  
coefficients.
(Note that because of the connection between type I (I') and type IIB  
(A) the
coupling $g$ that we have used should be identified with the type I  
(or I')
couplings $g,~(g')$.

Type I:
\begin{equation}\label{}
S=-{1\over (2\pi )^7}\int d^{10}x\sqrt G\left [{g^{-2}\over (\a
')^4}R+{g^{-1}\over 4(\a ')^3}\tr F^2+\ldots\right ]
\end{equation}

Putting $G_{\mu\nu}=g_H^{-1}G_{H\mu\nu}$and $g_H = 1/g$ we have,\\
Heterotic ($SO(32)$):
\begin{equation}\label{}
S=-{1\over (2\pi )^7}\int d^{10}x\sqrt {G_H}g_H^{-2}\left [{1\over  
(\a
')^4}R+{1\over 4(\a ')^3}\tr F^2+\ldots\right ]
\end{equation}
We also have (using (\ref{keleven}) in the expression for the action  
in
\cite{hw}),\\
M-theory on $ R_{10}\times S_1/Z_2$:
\begin{equation}\label{}
S=-{1\over (2\pi )^8(\a ')^{9/2}}\int d^{11}x\sqrt  
{G_M}R_M-\sum_i{1\over
4}{1\over (2\pi )^7(\a ')^3}\int_{M_i}\tr F_i^2+\ldots
\end{equation}
Transforming to the  heterotic metric
$ G_{Mmn}=g_H^{4/3}(dx^{11})^2+g_H^{-2/3}G_{\mu\nu}dx^{\mu}dx^{\nu}$,  
we
get,\\
Heterotic ($E_8\times E_8$):
\begin{equation}\label{}
S=-{1\over (2\pi )^7}\int d^{10}x\sqrt G{g}_H^{-2}\left [{1\over (\a
')^4}R+{1\over 4(\a ')^3}\sum_i\tr F_i^2+\ldots\right ]
\end{equation}

Finally putting $ G_{\mu\nu}=g_H{G'}_{\mu\nu},~g'=1/g_H$ we have,\\
Type I$'$:
\begin{equation}\label{}
S=-{1\over (2\pi )^7}\int d^{10}x\sqrt {G'}\left [{g'^{-2}\over (\a
')^4}R+{g'^{-1 }\over 4(\a ')^3}\sum_i\tr F_i^2+\ldots\right ]
\end{equation}

The point of the above expressions is that we have expressed the low  
energy
couplings in terms of the two physical quantities; the D-string  
tension or 11
dimensional physical radius, and the fundamental string tension.  We  
also see
that the value of the 10
dimensional type I' gauge coupling as determined in (\ref{gauge}) is  
consistent
with the Horava-Witten calculation \cite{hw} and heterotic-type I'  
duality.
Alternatively if we had assumed the latter duality we could have  
determined the
M-theory gauge coupling.

There is also another M theory number that can be fixed by our
arguments. This is the coefficient of the purely gravitational  
Green-Schwarz
term in the M theory action\footnote{I wish to thank E. Witten for  
suggesting
this check.} (equation (3.12) of  the second paper in \cite{hw} or  
equation
(3.14) of \cite{dlm} which was left undetermined in \cite{hw}). In
 \cite{dlm} this coefficient is given as the M theory membrane  
tension $T_2^M$.
Using our value
for this (\ref{twoM}) we get (note that in the expression below  $R$  
is the
curvature two form and $C_3$ is the three form field of M  
theory)\footnote{Our
conventions are the same as in \cite{dlm}. In particular our three  
form field
$C_{3}$ is related to that of \cite{hw} by $C_3=\sqrt 2 C_3^{HW}$. }

\begin{equation}\label{anom}
{1\over (2\pi)^2\a '^{3/2}}\int C_3\wedge{1\over (2\pi )^4}\left [  
-{1\over
768}(\tr R^2)^2+{1\over 192}\tr R^4 \right ].
\end{equation}
  On compactifying on a circle of coordinate radius $\sqrt{\a '}$  in
accordance with our convention (note that there are no metric factors   
in this
topological term), we get the correct expression with the right  
numerical
factor for the corresponding expression in the type IIA string (see  
\cite{dlm}
and references therein).  The coefficient above may be expressed in  
terms of
the 11D gravitational constant using (\ref{keleven}) and we find  
$[(2\pi
)^2/2\k^2_{11}]^{1/3} $. This gives a numerical coefficient which is  
a factor
of $1/3$ times the value given in \cite{hw}. At first sight this  
appears to
conflict with the requirement of anomaly cancellation, however it may  
be the
case that there is an alternative way of cancelling the anomalies in  
the theory
of \cite{hw}.\footnote{This matter is currently under investigation.}

To end this note let us  explore the consequences of the speculation    
that
M-theory/non-perturbative string theory effects actually pick the  
point
$g=g_H=g'={g'}_H=1$ since in some sense this is the most symmetrical  
choice.
The ten dimensional gravitational and gauge couplings are  then given  
by

\begin{equation}\label{}
2\k^2=(2\pi )^7\a '^4,~~\l^2=(2\pi )^7\a '^3
\end{equation}
giving the usual relation $2\k^2 =\a '\l^2 $ for heterotic strings
\cite{gross}, \cite{pg}. It should be noted that the argument in  
\cite{pg}
which depends only on the identification of the leading  singularity  
in the
operator product expansion of two vertex operators for gauge fields  
would be
independent of world sheet topology and hence should be valid at  
least to all
orders in perturbation theory.  The overall normalizations, which in  
the case
of weak coupling perturbation theory are given by $2\k^2=\shalf  
g^2(2\a
')^4,~\l^2=(2\a ')^3 $ \cite{gross}, are however changed.

Now Witten \cite{ew} has recently argued that strong coupling physics  
gives a
new perspective on the so-called string scale problem (For recent  
reviews see
\cite{kd}.). Let us see how this works out in our case. \footnote{  
For further
explorations of these phenomenological issues along the lines of  
\cite{ew}, see
\cite{bd},\cite{vk}.}
The four dimensional couplings are,

\begin{equation}\label{fourd}
G_N={(2\pi )^6\over 8}{\a '^4\over V},~~\a_{GUT}\equiv{\l_4^2\over
4\pi}=\shalf(2\pi )^6{\a '^3\over V}.
\end{equation}
Note that the gauge coupling can be small even though the string  
coupling is
unity, if the compactification scale $V^{1/6}$ is large compared to  
the natural
scale $2\pi\sqrt\a '$.
The relation between the gravitational and gauge  coupling constants  
is the
same as in weak
coupling (i.e. $G_N={1\over 4}\a_{GUT}\a '$). Nevertheless, because  
of the
change in the relation between $\a ' $ and $ \a_{GUT}$ (coming from  
the overall
factors of $2\pi$), we have a factor of $ \simeq 5$ improvement over  
the weak
coupling result.  Specifically, eliminating $\a '$ from (\ref{fourd})
we have $ G_N={\a_{GUT}^{4/3}(2V)^{1/3}/ 16\pi^2}$. If we assume that  
the
compactification scale  is given  by the `observed' unification  
scale, i.e.
$V^{-1/6}=M_{GUT}=2-3\times 10^{16} GeV$ and put  $\a_{GUT}^{-1}=25 $  
at that
scale \cite{ll}, we get a Planck mass of  $2-3\times 10^{18} GeV$,  
which is
still a factor of 4-5 too
small. However it is not clear what significance this has, since the
identification of the $V^{1/6}$ with the unification scale is just an  
order of
magnitude estimate and may be off by factors of $2\pi$ etc.. In any  
case there
is still the question of why the compactification volume is a factor  
10 larger
than the natural volume of string compactification $\simeq  
(2\pi\sqrt\a ')^6$
(which is what one has for four dimensional strings). Of course the  
length
scales  differ only by a factor of $\simeq1.5$!

\noindent{\bf Acknowledgments}\\
I wish to thank  Joe Polchinski and John Schwarz  for very useful  
discussions,
and E. Witten for a very useful e-mail message.
This work is
partially supported by the
Department of Energy contract No. DE-FG02-91-ER-40672.

\noindent{\bf Appendix}\\
\setcounter{equation}{0}
 In this appendix the 11D supergravity quantization condition of  
\cite{dlm} is
rederived and  shown to be in agreement with the (\ref{M}) and hence
(indirectly) also with (\ref{jp})\footnote{I wish to thank M.J. Duff  
and R.
Minasian for discussions.}. The argument
proceeds by using Dirac quantization at the membrane level and then  
at the 11D
effective action level.

Let us consider the integral of $K_4=dC_3=dC'_3$ over a 4-sphere  
where the two
guage fields are used respectively on   the upper and lower  
hemispheres.Then

\begin{eqnarray}
T_2\int_{S_4} K_4&=&T_2\int_{S^+_4} dC_3+T_2\int_{S^+_4} dC'_3\nn
&=& T_2\int_{S_3}(C_3-C_3')=2\pi n, ~~n\e{\cal Z}.
\end{eqnarray}
 The second equality follows from Stokes' theorem and the last from  
the
requirement that the membrane action should give a well defined  
quantum theory.
Let us now consider the manifold $M_{12} =S^4\times S^4\times S^4$.  
Using the
above result with $n=1$ we get,
\begin{equation}
{1\over 12\k_{11}^2}\int_{S^4\times S^4\times S^4}K_4\wedge K_4\wedge  
K_4
={3!\over 12\k_{11}^2}\left ({2\pi\over T_2}\right )^3
\end{equation}
On the other hand from Stokes' theorem, and the condition that the  
topological
term in the 11D supergravity action be quantum mechanically  
consistent, we get
for the left hand side of the above equation,
\begin{eqnarray}
{1\over 12\k_{11}^2}\left [\int_{M_{12}^+}dC_3\wedge K_4\wedge K_4
+\int_{M_{12}^-}dC'_3\wedge K_4\wedge K_4\right ] &=&{1\over
12\k_{11}^2}\int_{M_{11}}(C_3-C_3')\wedge K_4\wedge K_4\nn &=& 2\pi n
\end{eqnarray}
Comparing the two expressions we get

\begin{equation}
{(2\pi)^2\over 2\k_{11}^2 T_2^3}=m,~~m\e{\cal Z}.
\end{equation}
Combining this with the relation $2\k_{11}^2T_2 T_5=2\pi n$ which  
follows from
the membrane quantization condition and the existence of 5-branes, we  
get
$T_5/T_2^2 =n/(2\pi)$.

\end{document}